\documentclass[aps,prc,twocolumn,preprintnumbers,showpacs,floatfix,nofootinbib]{revtex4}
\usepackage{amsmath,graphicx}

\newcommand{\mean}[1]{\langle #1 \rangle} 

\begin{document}

\preprint{BI-TP 2006/47}

\title{Momentum conservation and correlation analyses in heavy-ion collisions 
  at ultrarelativistic energies}

\author{Nicolas Borghini}
\email{borghini@physik.uni-bielefeld.de}
\affiliation{Fakult\"at f\"ur Physik, Universit\"at Bielefeld,
Postfach 100131, D-33501 Bielefeld, Germany}
\date{\today}

\begin{abstract}
Global transverse-momentum conservation induces correlations between any number 
of particles, which contribute in particular to the two- and three-particle 
correlations measured in heavy-ion collisions. 
These correlations are examined in detail, and their importance for studies of 
jets and their interaction with the medium is discussed. 
\end{abstract}


\maketitle

\section{Introduction}

The high-statistics data collected at the Brookhaven Relativistic Heavy-Ion 
Collider (RHIC) have opened the possibility of novel precision measurements
aiming at characterizing the medium created in ultrarelativistic heavy-ion 
collisions. 
Among these new developments, significant effort is being devoted to extracting
correlations between pairs or triplets of high-momentum particles from the data,
to perform jet physics in the high-multiplicity environment of a high-energy 
nucleus--nucleus collision.
Thus, the initial result from a study of azimuthal correlations between 
particles with transverse momenta $2\mbox{ GeV}/c<p_T<6\mbox{ GeV}/c$ was the 
convincing demonstration that the back jet, at 180$^{\rm o}$ from a high-$p_T$
reference particle, is suppressed, in the sense that it does not emerge above 
the background~\cite{Adler:2002tq}.

More recently, the focus has been to obtain a more quantitative description of 
the behaviour in the recoil region away from a high-$p_T$ ``trigger'' particle.
On the one hand, various physical mechanisms have been proposed, involving the 
interaction of the high-momentum parton initially emitted back-to-back to the 
trigger with the medium through which it propagates, that predict non-trivial 
structures in this away-side region: 
shock-waves along a Mach cone~\cite{Baumgardt:1975qv} generated by the energy 
deposited by the fast-moving parton~\cite{Stoecker:2004qu,%
  Casalderrey-Solana:2004qm};
gluon bremsstrahlung~\cite{Vitev:2005yg} or Cherenkov-like radiation~\cite{%
  Dremin:2005an,Ruppert:2005uz,Koch:2005sx} by the parton;
the deflection of the back jet by the collective movement (``flow'') of the 
expanding medium~\cite{Armesto:2004pt};
or a path-length-based selection of the particles emerging a random walk 
through the medium~\cite{Chiu:2006pu}.

In parallel, several analysis techniques were developed and are being applied 
to the experimental data, using two-~\cite{Ajitanand:2005jj} or 
three-particle~\cite{Pruneau:2006gj,Ulery:2006iw,Ajitanand:2006is} correlations,
to distinguish between the different scenarios. 
A common need in these approaches is the necessity to deal with the background
properly. 
Thus, all measurements take into account the modulation of azimuthal 
correlations induced by anisotropic collective flow. 
However, the methods differ with respect to the other sources of correlations.
The philosophy of the cumulant study advocated in Ref.~\cite{Pruneau:2006gj} is 
to perform a measurement with minimal assumptions on the nature of the 
correlations, and later to compute the contributions (assumed to be additive) 
of the different sources of correlation to the three-particle cumulant. 
In contrast, all other methods rely on the assumption, be it explicitly stated 
or not, that the measured correlation consists of a jet signal, which vanishes 
away enough from the jet, and an uncorrelated background. 

The purpose of this paper is to emphasize the role of the unavoidable 
contribution of {\em global\/} transverse momentum conservation to the measured 
correlation, either between two or three particles. 
The realization that global momentum conservation can significantly affect 
measurements in heavy-ion physics is not new. 
Thus early measurements of collective anisotropic flow in nuclear collisions at 
intermediate and relativistic energies were corrected~\cite{Danielewicz:1988in} 
for that effect.
Later, it was shown that momentum conservation biases the standard measurement 
of the first anisotropic-flow harmonic $v_1$ at SPS~\cite{Borghini:2000cm} and 
an explicit correction to the analysis was devised~\cite{Borghini:2002mv} and 
applied to the NA49 data~\cite{Alt:2003ab}.
Recently, the possibility that the conservation of momentum and energy could 
significantly affect the correlation functions measured in femtoscopy analyses 
was addressed in Ref.~\cite{Chajecki:2006hn}.

Here I shall argue that correlations at high transverse momentum are especially 
sensitive to the conservation of total transverse momentum. 
The latter has several effects. 
On the one hand, it contributes to the cumulant measured in the technique of 
Ref.~\cite{Pruneau:2006gj}.
The specific dependence on the particle momenta of the three-particle cumulant 
from global transverse momentum conservation, which was already computed in 
Ref.~\cite{Borghini:2003ur}, will thus be investigated in Sec.~\ref{s:3p-cum}.
On the other hand, global momentum conservation correlates a jet to the 
remainder of the event, thereby invalidating the assumption that jet and 
background are uncorrelated. 
Some implications of this observation will be discussed in 
Sec.~\ref{s:comments}.

\section{Momentum conservation and cumulants}
\label{s:3p-cum}

In a nucleus--nucleus collision, all $N$ emitted particles are correlated 
together by the requirement that their transverse momenta ${{\bf p}_T}_i$ add up
to ${\bf 0}$. 
As a consequence, the joint $M$-particle probability distribution 
$f({{\bf p}_T}_{j_1},\ldots,{{\bf p}_T}_{j_M})$ of ${{\bf p}_T}_{j_1}$, \ldots, 
${{\bf p}_T}_{j_M}$ differs from the product of the single-particle probability 
distributions $f({{\bf p}_T}_{j_1}) \cdots f({{\bf p}_T}_{j_M})$, where 
$1\leq j_1 < \ldots < j_M\leq N$ and $2\leq M\leq N$.
In particular, the cumulant of the $M$-particle distribution, which will be 
denoted by $f_c({{\bf p}_T}_{j_1},\ldots,{{\bf p}_T}_{M_1})$ and corresponds to 
the part of the joint probability distribution that cannot be expressed in 
terms of distributions involving less than $M$ particles~\cite{vanKampen}, is 
finite. 
Using a generating function of the multiparticle probability distributions and 
a saddle-point calculation, one can compute the successive cumulants to leading 
order in $1/N$~\cite{Borghini:2003ur}. 
Thus, it was shown that the $M$-particle cumulant scales like $1/N^{M-1}$. 
Assuming that the single-particle ${\bf p}_T$ distribution is isotropic, one in 
particular finds that the two- and three-particle cumulants due to global 
transverse-momentum conservation read
\begin{widetext}
\begin{eqnarray}
\label{2p-cumulant}
f_c({{\bf p}_T}_1, {{\bf p}_T}_2) &=& \displaystyle 
-\frac{2\,{{\bf p}_T}_1\cdot {{\bf p}_T}_2}{N \mean{{\bf p}_T^2}}, \\
f_c({{\bf p}_T}_1, {{\bf p}_T}_2, {{\bf p}_T}_3) & = & \displaystyle
-\frac{2}{N^2\mean{{\bf p}_T^2}} 
({{\bf p}_T}_1 \cdot {{\bf p}_T}_2 + {{\bf p}_T}_1 \cdot {{\bf p}_T}_3 + {{\bf p}_T}_2 \cdot {{\bf p}_T}_3) \cr
 & & + \displaystyle \frac{4}{N^2\mean{{\bf p}_T^2}^2} \left[
({{\bf p}_T}_1 \cdot {{\bf p}_T}_2)({{\bf p}_T}_1 \cdot {{\bf p}_T}_3) + 
({{\bf p}_T}_1 \cdot {{\bf p}_T}_2)({{\bf p}_T}_2 \cdot {{\bf p}_T}_3) + 
({{\bf p}_T}_1 \cdot {{\bf p}_T}_3)({{\bf p}_T}_2 \cdot {{\bf p}_T}_3) \right], 
\label{3p-cumulant}
\end{eqnarray}
\end{widetext}
where $\mean{{\bf p}_T^2}$ denotes the average over many particles and events of
the squared transverse momentum.

The meaning of the two-particle cumulant~(\ref{2p-cumulant}) is clear and 
intuitive: the correlation is back-to-back and stronger between particles with 
higher transverse momenta. 
In plain words, given a high-$p_T$ trigger-particle, there is a larger 
probability to find an ``associated'' particle away from it in azimuth than 
close to it, just because of global transverse-momentum conservation. 

The interpretation of the three-particle cumulant arising from global 
transverse-momentum conservation, Eq.~(\ref{3p-cumulant}), is slightly more 
involved, as it implies two terms with opposite signs. 
Yet, if one considers three particles with transverse momenta significantly 
larger than the rms transverse momentum $\mean{{\bf p}_T^2}^{1/2}$, one sees that
the ``attractive'' term, in the second line, dominates over the ``repulsive'' 
one.
To illustrate the behaviour of the three-particle cumulant, I shall now study 
Eq.~(\ref{3p-cumulant}) by promoting particle 1 to the role of ``trigger 
particle,'' with respect to which the azimuths of the other two are measured:
$\Delta\varphi_{12}\equiv\varphi_2-\varphi_1$, 
$\Delta\varphi_{13}\equiv\varphi_3-\varphi_1$. 
I shall assume for simplicity that the ``associated particles'' 2 and 3 have 
equal transverse momenta ${p_T}_2\!={p_T}_3\leq {p_T}_1$, and use the notation
$f_c({p_T}_1,{p_T}_2,{p_T}_3,\Delta\varphi_{12},\Delta\varphi_{13})$ for the 
cumulant. 

Consider first the symmetric case $\Delta\varphi_{13}=2\pi-\Delta\varphi_{12}$. 
For symmetry reasons (parity and $2\pi$-periodicity), the three-particle 
cumulant $f_c({p_T}_1,{p_T}_2,{p_T}_2,\Delta\varphi_{12},-\Delta\varphi_{12})$ has 
trivial extrema at $\Delta\varphi_{12}=0$ and $\Delta\varphi_{12}=\pi$. 
The former is a maximum for values of ${p_T}_2\!={p_T}_3$ larger than a minimal 
value
\begin{eqnarray*}
{p_T}_{\rm min}\!&\equiv& \frac{\mean{{\bf p}_T^2}}{10\,{p_T}_1}\left(
  \frac{-{p_T}_1^2}{\mean{{\bf p}_T^2}}+1 + 
  \sqrt{\frac{{p_T}_1^4}{\mean{{\bf p}_T^2}^2}+
    \frac{8{p_T}_1^2}{\mean{{\bf p}_T^2}}+1} \right) \\
 &\simeq& \frac{\mean{{\bf p}_T^2}}{2{p_T}_1},
\end{eqnarray*}
i.e., for practical purposes in high-$p_T$ studies, always, since 
${p_T}_1\gg\mean{{\bf p}_T^2}^{1/2}$ (as has been assumed in the second line of 
the above identity).
For values of the associated-particle transverse momenta ${p_T}_2\!={p_T}_3$ 
larger than the same ${p_T}_{\rm min}$, the three-particle cumulant $f_c$ also has
a minimum for 
\begin{eqnarray*}
\lefteqn{\cos\Delta\varphi_{12}=} & \\
 & \displaystyle\frac{\mean{{\bf p}_T^2}}{12\,{p_T}_1{p_T}_2}\!\!\left[
  1\!-\!\frac{{p_T}_1^2}{\mean{{\bf p}_T^2}} \!+\! 
  \sqrt{\!\left(\!\frac{{p_T}_1^2}{\mean{{\bf p}_T^2}}\!-\!1\!\right)^{\!\!2}\!\!+
    \!\frac{12{p_T}_1^2}{\mean{{\bf p}_T^2}}\!\!\left(\!
      1\!+\!\frac{2{p_T}_2^2}{\mean{{\bf p}_T^2}}\right)\!} \right]\!.
\end{eqnarray*}
One can check that the $\Delta\varphi_{12}$ position of this minimum is an 
increasing function of ${p_T}_2$, which  reaches a maximal value 
$\Delta\varphi_{12}=\arccos(1/3 + \mean{{\bf p}_T^2}/6{p_T}_1^2)\simeq 1.23$~rad 
(70.5$^{\rm o}$) for ${p_T}_2\!={p_T}_1$. 
The next extremum of the three-particle cumulant in the range 
$0\leq\Delta\varphi_{12}\leq\pi$ only exists if the transverse momentum of the 
associated particles ${p_T}_2$ is larger than 
\[
{p'_T}_{\rm min}\!\equiv \frac{\mean{{\bf p}_T^2}}{10\,{p_T}_1}\left(
  \frac{{p_T}_1^2}{\mean{{\bf p}_T^2}}-1 + 
  \sqrt{\frac{{p_T}_1^4}{\mean{{\bf p}_T^2}^2}+
    \frac{8{p_T}_1^2}{\mean{{\bf p}_T^2}}+1} \right), 
\]
which is $\simeq {p_T}_1/5$ when the transverse momentum ${p_T}_1$ of the trigger
particle is large. 
Then there is a second local maximum, beyond that at $\Delta\varphi_{12}=0$, for
\begin{eqnarray*}
\lefteqn{\cos\Delta\varphi_{12}=} & \\
 & \displaystyle\frac{\mean{{\bf p}_T^2}}{12\,{p_T}_1{p_T}_2}\!\!\left[
  1\!-\!\frac{{p_T}_1^2}{\mean{{\bf p}_T^2}} \!-\! 
  \sqrt{\!\left(\!\frac{{p_T}_1^2}{\mean{{\bf p}_T^2}}\!-\!1\!\right)^{\!\!2}\!\!+
    \!\frac{12{p_T}_1^2}{\mean{{\bf p}_T^2}}\!\!\left(\!
      1\!+\!\frac{2{p_T}_2^2}{\mean{{\bf p}_T^2}}\right)\!} \right]\!.
\end{eqnarray*}
The position of this maximum decreases with increasing ${p_T}_2$, reaching a 
minimal $\Delta\varphi_{12}=2\pi/3$ when ${p_T}_2\!={p_T}_1$. 
\begin{figure*}[t!]
\includegraphics[width=0.49\linewidth]{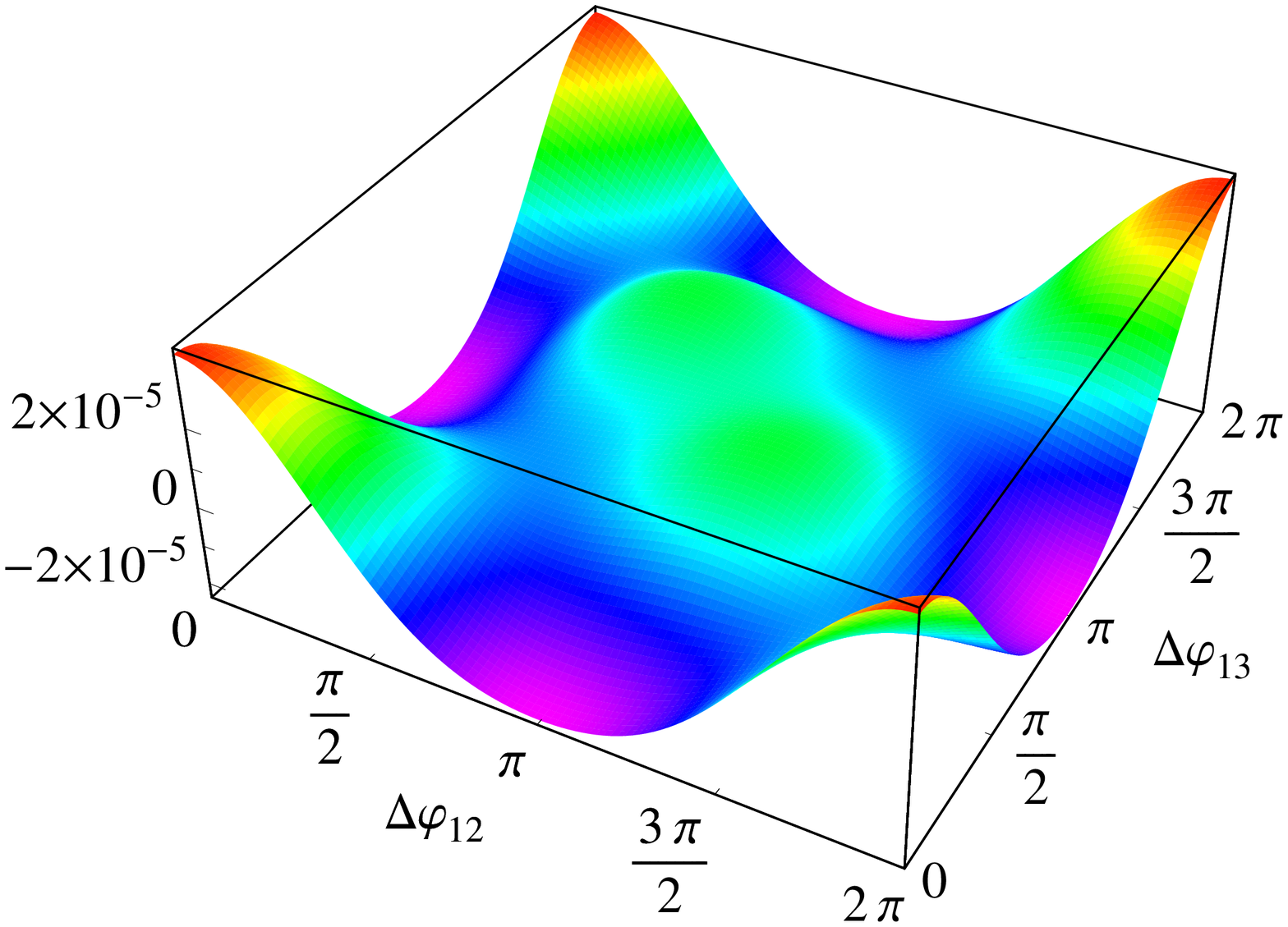}\hfill%
\includegraphics[width=0.49\linewidth]{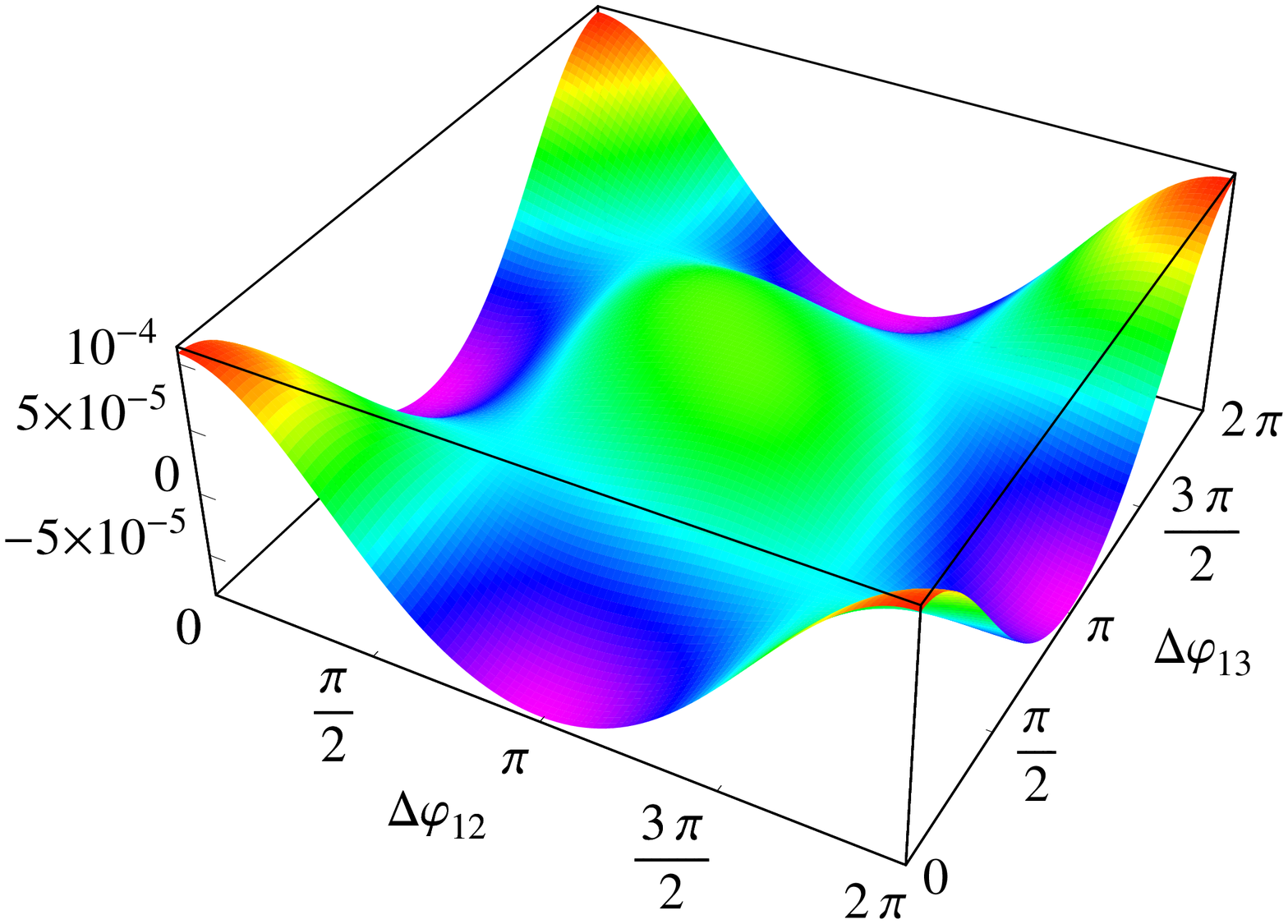}\\[-3mm]
\caption{(Color online) Three-particle cumulant due to global 
  transverse-momentum conservation 
  $f_c({p_T}_1,{p_T}_2,{p_T}_3,\Delta\varphi_{12},\Delta\varphi_{13})$, as a 
  function of the relative angles $\Delta\varphi_{12}$, $\Delta\varphi_{13}$, 
  assuming $N=8000$ particles per event with a rms transverse momentum 
  $\mean{{\bf p}_T^2}^{1/2}=0.45\mbox{ GeV}/c$. 
  Left: ${p_T}_1\!=3.2\mbox{ GeV}/c$, ${p_T}_2\!={p_T}_3\!=1.2\mbox{ GeV}/c$; 
  right: ${p_T}_1\!=6\mbox{ GeV}/c$, ${p_T}_2\!={p_T}_3\!=1.2\mbox{ GeV}/c$.}
\label{fig:3p-cum}
\end{figure*}
This is actually a rather intuitive result: when the three particles have equal 
transverse momenta, the cumulant has a maximum at the symmetric configuration 
$\Delta\varphi_{12}=\Delta\varphi_{23}=\Delta\varphi_{31}=120^{\rm o}$.
Eventually, the three-particle cumulant 
$f_c({p_T}_1,{p_T}_2,{p_T}_2,\Delta\varphi_{12},-\Delta\varphi_{12})$ has the 
already-mentioned extremum at $\Delta\varphi_{12}=\pi$, which is a minimum if 
${p_T}_2>{p'_T}_{\rm min}$, a maximum otherwise. 

Let me now study the three-particle cumulant $f_c$ in the specific case 
${p_T}_2\!={p_T}_3$ and $\Delta\varphi_{13}=\Delta\varphi_{12}$. 
Due to parity and $2\pi$-periodicity, it has two extrema at 
$\Delta\varphi_{12}=0$ and $\Delta\varphi_{12}=\pi$. 
The latter is always a maximum, irrespective of the value of ${p_T}_2$. 
That means that the point $\Delta\varphi_{12}=\Delta\varphi_{13}=\pi$ is either a
saddle point for the cumulant
$f_c({p_T}_1,{p_T}_2,{p_T}_2,\Delta\varphi_{12},\Delta\varphi_{13})$, if 
${p_T}_2>{p'_T}_{\rm min}$, or a local maximum if ${p_T}_2\leq {p'_T}_{\rm min}$.
The nature of the extremum at $\Delta\varphi_{12}=0$ depends on the value of the 
transverse momentum of the associated particle. 
If ${p_T}_2\leq (\sqrt{{p_T}_1^2-2\mean{{\bf p}_T^2}}-{p_T}_1)/2$ --- which is 
$\simeq\mean{{\bf p}_T^2}/2{p_T}_1\ll\mean{{\bf p}_T^2}^{1/2}$ in the regime 
${p_T}_1\gg\mean{{\bf p}_T^2}^{1/2}$, so that the case is most likely irrelevant 
for studies of correlations at high transverse momentum --- then 
the cumulant has a minimum at $\Delta\varphi_{12}=0$. 
On the contrary, when ${p_T}_2>(\sqrt{{p_T}_1^2-2\mean{{\bf p}_T^2}}-{p_T}_1)/2$
the cumulant has a maximum at $\Delta\varphi_{12}=0$, and a minimum for
\[
\cos\Delta\varphi_{12}=
-\frac{2{p_T}_2^2-\mean{{\bf p}_T^2}}{2{p_T}_1{p_T}_2}.
\]
The position of this minimum grows with ${p_T}_2$, reaching 
$\Delta\varphi_{12}=\arccos(-1+\mean{{\bf p}_T^2}/2{p_T}_1^2)$ for 
${p_T}_2\!={p_T}_1$. 
In the case of a large transverse momentum ${p_T}_1$ of the trigger particle, 
this minimum sits at 
$\Delta\varphi_{12}\simeq\pi-\sqrt{\mean{{\bf p}_T^2}}/\sqrt{2}{p_T}_1$, close to 
the maximum at $\pi$. 

Figure~\ref{fig:3p-cum} shows the profile of the three-particle cumulant due to 
global transverse-momentum conservation 
$f_c({p_T}_1,{p_T}_2,{p_T}_3,\Delta\varphi_{12},\Delta\varphi_{13})$, in the case 
of equal transverse momenta of the associated particles, ${p_T}_2\!={p_T}_3$, 
larger than the rms transverse momentum $\mean{{\bf p}_T^2}^{1/2}$. 
Both choices ${p_T}_2\!>{p'_T}_{\rm min}\!\sim{p_T}_1/5$ (left) and 
${p_T}_2\!<{p'_T}_{\rm min}$ (right) are displayed to illustrate the different 
behaviours discussed above. 
As anticipated, in the first case, the cumulant has a saddle point at 
$\Delta\varphi_{12}=\Delta\varphi_{13}=\pi$ and two clearly separated maxima away 
from $\Delta\varphi_{12}=\Delta\varphi_{13}=0$ (modulo $2\pi$) along the line 
$\Delta\varphi_{12}=2\pi-\Delta\varphi_{13}$; whereas in the second case, there 
are two maxima for values of $\Delta\varphi_{12}=\Delta\varphi_{13}$ equal to 0 
or $\pi$, and no further structure. 

The values of the total number of particles $N=8000$ and the rms transverse
momentum $\mean{{\bf p}_T^2}^{1/2}=450\mbox{ MeV}/c$ adopted in 
Fig.~\ref{fig:3p-cum} were chosen so as to mimic a central Au--Au collision at 
RHIC. 
Such numbers result in a three-particle cumulant from transverse-momentum 
conservation of order $10^{-5}$ for transverse momenta of the trigger and 
associated particles of 3 and 1 GeV/$c$ respectively. 
This is admittedly a small correlation, yet it is of the same size as the 
contribution of anisotropic collective flow to the cumulant~\cite{%
  Pruneau:2006gj}. 
In addition, the correlation will be stronger in more peripheral collisions, 
since it scales as the inverse squared multiplicity $1/N^2$, see 
Eq.~(\ref{3p-cumulant}).\footnote{The rms transverse momentum 
  $\mean{{\bf p}_T^2}^{1/2}$ may also decrease when going to more peripheral 
  collisions, which contributes, although much less than the drop in 
  multiplicity, to the growth of the cumulant.}
Similarly, the value of the correlation will be larger in smaller systems and 
at lower collision energies. 
In fact, the multiplicity $N$ that enters Eqs.~(\ref{2p-cumulant}-\ref{%
  3p-cumulant}) might not be the {\em total\/} event multiplicity, but one could
argue that different rapidity slices are decorrelated, so that the constraint 
from transverse momentum conservation would actually be driven by a smaller 
multiplicity. 
That would also give values of the cumulant of the correlation from momentum 
conservation higher than those plotted in Fig.~\ref{fig:3p-cum}.

\section{Discussion}
\label{s:comments}

Till now, I have mostly focussed on the cumulant~(\ref{3p-cumulant}) of the 
three-particle correlation due to global transverse-momentum conservation. 
\begin{figure}[t!]
\centerline{\includegraphics[width=0.98\linewidth]{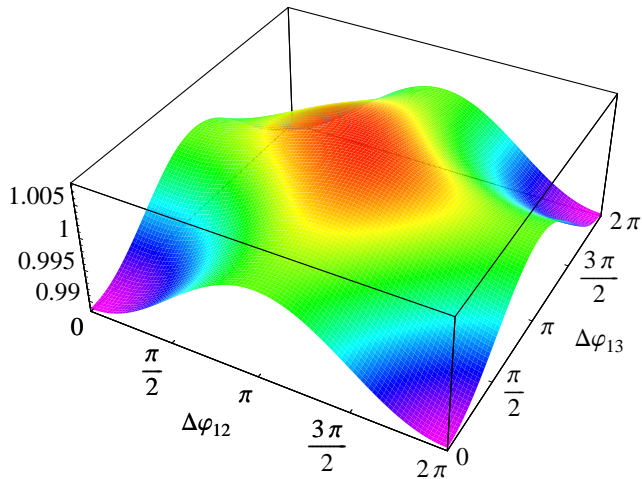}}
\caption{(Color online) Three-particle correlation due to global 
  transverse-momentum conservation vs.\ the relative angles $\Delta\varphi_{12}$,
  $\Delta\varphi_{13}$, assuming $N=8000$ particles per event with a rms 
  transverse momentum $\mean{{\bf p}_T^2}^{1/2}=0.45\mbox{ GeV}/c$, for particles
  of transverse momentum ${p_T}_1\!=3.2\mbox{ GeV}/c$ and 
  ${p_T}_2\!={p_T}_3\!=1.2\mbox{ GeV}/c$, respectively.}
\label{fig:3p-cor}
\end{figure}
The three-particle correlation itself, which also involves two-body terms 
[controlled by Eq.~(\ref{2p-cumulant})], is much larger, as displayed in 
Fig.~\ref{fig:3p-cor} for the same RHIC-inspired values of the multiplicity, 
rms transverse momentum, and transverse momenta of the trigger and associated 
particles as in the left panel of Fig.~\ref{fig:3p-cum}. 
One sees that the main, quite intuitive effect of momentum conservation is to 
push the associated particles 2 and 3 away from the trigger particle 1. 
The resulting ``bump'' at $\Delta\varphi_{12}=\Delta\varphi_{13}=\pi$ is broader 
along the $\Delta\varphi_{12}=2\pi-\Delta\varphi_{13}$ axis than along the 
perpendicular $\Delta\varphi_{12}=\Delta\varphi_{13}$ axis. 
This is quite generic for the correlation due to momentum conservation --- 
configurations with the two associated particles on each side of the direction 
defined by ${{\bf p}_T}_1$ are more likely than those with both particles on 
the same side ---, while the symmetric shape of the bump across the 
$\Delta\varphi_{12}=\Delta\varphi_{13}$ axis in Fig.~\ref{fig:3p-cor} reflects
the specific choice ${p_T}_2\!={p_T}_3$.

A most prominent feature of the profile, shown in Fig.~\ref{fig:3p-cor}, of the 
three-particle correlation arising from the momentum-conservation constraint, 
is the ``dip'' around the origin $\Delta\varphi_{12}=\Delta\varphi_{13}=0$. 
The meaning of this dip is transparent: if there is a high-$p_T$ particle in 
the event, then its transverse momentum has to be balanced by the others.
Therefore, the probability of finding another high-$p_T$ particle pointing into 
the same direction is smaller than if transverse momentum were not conserved.%
\footnote{Surprisingly, this mutual exclusion of high transverse-momentum 
  particles seems to be entirely driven by the anticorrelation in the two-body 
  term. 
  On the contrary, unless the associated particles have transverse momenta 
  smaller than $\mean{{\bf p}_T^2}^{1/2}$, the three-particle cumulant is 
  maximal at $\Delta\varphi_{12}=\Delta\varphi_{13}=0$: at that level, all 
  three high-$p_T$ particles are grouped together by the attractive term in 
  the second line of Eq.~(\ref{3p-cumulant}), rather than oriented away from 
  each other.}
In other words, a jet and the remainder of the ``underlying'' event are not 
uncorrelated, as is too often assumed, but global transverse-momentum 
conservation alone already constrains the momentum distribution of the 
particles ``outside the jet.'' 

A consequence of the existence of this correlation is that the mere notion of 
there being ``an underlying event over which the jet develops'' is incorrect. 
The jet does distort the event to which it belongs, it is not merely embedded 
in it as is done in many simulations. 
This means that, at least in azimuth, there is no ``region far from the jet'' 
where its influence would vanish, thereby allowing one to determine a correctly
normalized ``background'' to the jet: even in the over-simplified case of a 
one-particle jet and considering only two-particle correlations, the region 
where the event shape is not modified by the presence of the jet is restricted 
to two points only, at 90$^{\rm o}$ away from it. 
Even in studies of correlations between ``only'' two particles, the approaches 
currently used to disentangle the ``jet'' from the harmonic modulation of 
azimuthal correlations due to anisotropic flow may thus be inaccurate, since 
they ignore the modulation induced by transverse-momentum conservation.
The effect of the latter will be even more important in three-particle 
correlation studies. 
\medskip

In summary, I have shown that the conservation of the total transverse momentum 
yields a correlation between pairs or triplets of high-$p_T$ particles that is 
sizable --- in central Au--Au collisions at RHIC energies the resulting 
three-particle cumulant is of the same magnitude as that due to anisotropic 
collective flow --- and therefore should be accounted for properly in 
experimental studies. 
This might be easily feasible in studies at the cumulant level~\cite{%
  Pruneau:2006gj}: there one only has to consider the three-particle 
cumulant~(\ref{3p-cumulant}) --- possibly generalized to include the known
anisotropy of the ${\bf p}_T$ distribution (see Ref.~\cite{Borghini:2003ur}), 
especially in non-central collisions ---, so that the non-measured multiplicity 
$N$ and rms transverse momentum $\mean{{\bf p}_T^2}^{1/2}$ enter the analysis 
only once. 
The three-particle cumulant from transverse-momentum conservation has a 
specific dependence on the values of the particle momenta, which was illustrated
in special cases in Fig.~\ref{fig:3p-cum}. 
If the transverse momentum of the trigger particle is much larger (by a factor 
$\sim 5$) than those of the associated particles, the cumulant has a simple 
shape with one peak on the trigger-particle side, and a broader back-to-back 
bump, more elongated along the $\Delta\varphi_{12}=2\pi-\Delta\varphi_{13}$ axis 
than perpendicular to it (see Fig.~\ref{fig:3p-cum}, right panel). 
However, if one decreases the trigger transverse momentum while keeping fixed 
the associated transverse momenta, then a two-bump structure develops along the 
$\Delta\varphi_{12}=2\pi-\Delta\varphi_{13}$ axis on the side away from the 
trigger (Fig.~\ref{fig:3p-cum}, left).

Taking properly into account the effect of total momentum conservation will be 
much more involved in analyses that rely on an assumed jet profile~\cite{%
  Ajitanand:2005jj,Ulery:2006iw,Ajitanand:2006is,Kniege@QM06}. 
In those, the effect of momentum conservation also has to be considered at the 
two-body level, in the correlations between ``jet'' and ``background'' or 
between two ``background'' particles (although the correlation is smaller 
between softer particles). 
In addition, the correlation induced by momentum conservation somehow blurs the 
distinction between jet and background, since the latter is unquestionably 
different in the presence of the jet from what it would be in its absence. 
Before trying to identify definite structures, which require high-precision 
measurements, in the recoil region of a high-$p_T$ particle, it is important to 
first determine precisely the reference pattern --- including the effects of 
momentum conservation and anisotropic collective flow --- over which they would 
develop.


\begin{thebibliography}{99}

\bibitem{Adler:2002tq}
C.~Adler {\it et al.}  [STAR Collaboration],
Phys.\ Rev.\ Lett.\  {\bf 90}, 082302 (2003). 

\bibitem{Baumgardt:1975qv}
H.~G.~Baumgardt {\it et al.}, Z.\ Phys.\ A {\bf 273}, 359 (1975).

\bibitem{Stoecker:2004qu}
H.~St\"ocker, Nucl.\ Phys.\ A {\bf 750}, 121 (2005). 

\bibitem{Casalderrey-Solana:2004qm}
J.~Casalderrey-Solana, E.~V.~Shuryak, and D.~Teaney,
J.\ Phys.\ Conf.\ Ser.\ {\bf 27}, 22 (2005). 

\bibitem{Vitev:2005yg}
I.~Vitev, Phys.\ Lett.\ B {\bf 630}, 78 (2005). 

\bibitem{Dremin:2005an}
I.~M.~Dremin, Nucl.\ Phys.\ A {\bf 767}, 233 (2006). 

\bibitem{Ruppert:2005uz}
J.~Ruppert and B.~M\"uller, Phys.\ Lett.\ B {\bf 618}, 123 (2005). 

\bibitem{Koch:2005sx}
V.~Koch, A.~Majumder, and X.~N.~Wang, 
Phys.\ Rev.\ Lett.\ {\bf 96}, 172302 (2006). 

\bibitem{Armesto:2004pt}
N.~Armesto, C.~A.~Salgado, and U.~A.~Wiedemann,
Phys.\ Rev.\ Lett.\ {\bf 93}, 242301 (2004). 

\bibitem{Chiu:2006pu}
C.~B.~Chiu and R.~C.~Hwa, Phys.\ Rev.\ C {\bf 74}, 064909 (2006).

\bibitem{Ajitanand:2005jj}
N.~N.~Ajitanand {\it et al.}, Phys.\ Rev.\ C {\bf 72}, 011902 (2005). 

\bibitem{Pruneau:2006gj}
C.~A.~Pruneau, Phys.\ Rev.\ C {\bf 74}, 064910 (2006).

\bibitem{Ulery:2006iw}
J.~G.~Ulery and F.~Wang, nucl-ex/0609016;
nucl-ex/0609017.

\bibitem{Ajitanand:2006is}
N.~N.~Ajitanand, nucl-ex/0609038.

\bibitem{Danielewicz:1988in}
P.~Danielewicz {\it et al.}, Phys.\ Rev.\ C {\bf 38}, 120 (1988).

\bibitem{Borghini:2000cm}
N.~Borghini, P.~M.~Dinh, and J.-Y.~Ollitrault,
Phys.\ Rev.\ C {\bf 62}, 034902 (2000). 

\bibitem{Borghini:2002mv}
N.~Borghini {\em et al.\/}, Phys.\ Rev.\ C {\bf 66}, 014901 (2002). 

\bibitem{Alt:2003ab}
C.~Alt {\it et al.}  [NA49 Collaboration],
Phys.\ Rev.\ C {\bf 68}, 034903 (2003). 

\bibitem{Chajecki:2006hn}
Z.~Chaj\c ecki and M.~Lisa, nucl-th/0612080.

\bibitem{Borghini:2003ur}
N.~Borghini, Eur.\ Phys.\ J.\ C {\bf 30}, 381 (2003). 

\bibitem{vanKampen}
N.~G.~van Kampen, 
{\em Stochastic Processes in Physics and Chemistry}
(North Holland, Amsterdam, 1981).

\bibitem{Kniege@QM06}
S.~Kniege, talk given at Quark Matter 2006. 

\end{thebibliography}
\end{document}